\documentclass[twocolumn,amsmath,amssymb]{jpsj3}
\usepackage{color}
\usepackage{ulem}
\usepackage{float}

\usepackage{bm}
\usepackage{amsmath,amsfonts,amsthm,amssymb,slashbox}

\newcommand{\sP}{\mathcal{P}}

\newcommand{\ga}{\alpha}

\newcommand{\vr}{{\bm{r}}}

\def\journal#1#2#3#4#5{#1: #2 {\bf #3} (#5) #4.}

\def\JPSJ{J.\ Phys.\ Soc.\ Jpn.}
\def\PR{Phys.\ Rev.}
\def\PRL{Phys.\ Rev.\ Lett.}
\def\PRB{Phys.\ Rev.\ B}

\newcommand{\lU}{\ensuremath{\lambda_U}}

\newcommand{\lV}{\ensuremath{\lambda_{V,J}}}
\newcommand{\lc}{\ensuremath{\lambda_\mathrm{c}}}
\newcommand{\ms}{\ensuremath{m_\mathrm{s}}}

\newcommand{\Ns}{\ensuremath{N_\mathrm{s}}}

\definecolor{darkgreen}{rgb}{0.0, 0.7, 0.0} 


\def\runauthor{\textsc{H. Shinaoka} \textit{et al.}}
\makeatletter
\def\@typeset{}
\makeatother

\begin{document}

\title{Mott Transition and Phase Diagram of $\kappa$-(BEDT-TTF)$_2$Cu(NCS)$_2$ Studied by Two-Dimensional Model Derived from Ab initio Method}
\author{Hiroshi {\sc Shinaoka}$^{1,2}$\thanks{E-mail address: h.shinaoka@aist.go.jp}, Takahiro {\sc Misawa}$^{2,3}$, Kazuma {\sc Nakamura}$^{2,3}$, and Masatoshi {\sc Imada}$^{2,3}$}
\inst{
$^{1}$ Nanosystem Research Institute, AIST, Tsukuba 305-8568\\
$^{2}$ CREST, JST, 7-3-1 Hongo, Bunkyo-ku, Tokyo 113-8656\\
$^{3}$ Department of Applied Physics, University of Tokyo, Hongo, Bunkyo-ku, Tokyo 113-8656} 
\recdate{\today}

\date{\today}
\abst{
We present an {\it ab initio} analysis for the ground-state properties of a correlated organic compound $\kappa$-(BEDT-TTF)$_2$Cu(NCS)$_2$.
First, we derive an effective two-dimensional low-energy model from first principles, having short-ranged transfers and short-ranged Coulomb and exchange interactions.
Then, we perform many-variable variational Monte Carlo calculations for this model and draw a ground-state phase diagram as functions of scaling parameters for the onsite and off-site interactions.
The phase diagram consists of three phases; a paramagnetic metallic phase, an antiferromagnetic (Mott) insulating phase, and a charge-ordered phase.
In the phase diagram, the parameters for the real compound are close to the first-order Mott transition, being consistent with experiments.
We show that the off-site Coulomb and exchange interactions affect the phase boundary; (i) they appreciably stabilize the metallic state against the Mott insulating phase and 
(ii) enhance charge fluctuations in a wide parameter region in the metallic phase.
 We observe arc-like structure in Fermi surface around the region where the charge fluctuations are enhanced.
Possible relevance of the charge fluctuations to the experimentally observed dielectric anomaly in the $\kappa$-BEDT-TTF family compounds is also pointed out.}
\kword{organic conductors, first principles, Hubbard-type low-energy model, variational Monte Carlo method, Mott transition}

\maketitle

\section{Introduction}
In organic conductors, we find diverse properties and a plenty of phases including normal metals, superconductors and various types of insulators with antiferromagnetic, charge or (spin) Peierls orders as well as spin-liquid-type nonmagnetic Mott insulators.~\cite{Kanoda} Although the unit cells of these conductors contain many atoms constituting the molecules with complicated crystal structures, band structures near the Fermi level are in most cases simple with a small number of bands isolated from other bands located away from the Fermi level. These isolated bands originate from molecular orbitals [lowest unoccupied molecular orbitals (LUMO) and highest occupied molecular orbitals (HOMO)]. Because the number of bands near the Fermi level is small, screening of the electron-electron Coulomb interaction is poor. In addition and more importantly, large lattice constants make the overlap of the neighboring molecular orbitals small, leading to a large ratio of the screened electron interaction to the kinetic energy.  This is the reason why electron correlations are in general strong in the organic conductors.  
Because of the complex unit cell and prominent strong correlation effects, the {\it ab initio} calculation and clarification of mechanisms of material properties in the organic conductors remain as big challenges. 

Among all, a family of compounds, (ET)$_2X$ with a number of choices of anions $X$ alternatingly stacked with BEDT-TTF molecules [where BEDT-TTF is bis(ethylenedithio)-tetrathiafulvalene, abbreviated as ET], offers a variety of prototypical behaviors of strongly correlated electron systems with two-dimensional (2D) anisotropies.~\cite{Kanoda} 
In particular, the $\kappa$-type compounds characterized by dimerization of ET molecules have served to discoveries of unconventional quantum phases and unexplored concepts at the forefront of condensed matter physics. 

Unconventional superconductivity is found in some of these compounds that show metallic behaviors.  Namely, the compounds with the anions $X$=Cu[N(CN)$_2$]Br (ref.~\citen{Geiser}) and $X$=Cu(NCS)$_2$ (refs.~\citen{CuSCN,Schultz}) abbreviated as $\kappa$-Br and $\kappa$-NCS hereafter, respectively, show superconducting transitions at $T_c\sim$10-13K. 
Under pressure, the compound with $X=$Cu$_2$(CN)$_3$  referred to as $\kappa$-CN also shows superconductivity below 2.8 K.~\cite{Cu2CN3} However, the driving mechanism of the superconductivity is not fully understood yet.  

An unconventional nonmagnetic Mott-insulating phase found near the Mott transition for $\kappa$-CN is another example of such a discovery. In contrast to a naive expectation for a magnetic order at sufficiently low temperatures, no apparent magnetic orders are observed even at a prominently low temperature $T$=0.03 K that is four orders of magnitude lower than the antiferromagnetic spin-exchange interaction $J$$\sim$250 K.~\cite{Shimizu} 

The emergence of the quantum spin liquid near the Mott transition on two types of 2D Hubbard models with geometrical frustration effects has already been predicted in earlier numerical studies~\cite{Kashima,Morita,Mizusaki} with the help of essentially exact algorithm of the path integral renormalization group (PIRG).~\cite{Imada2000,Kashima2001} The ground state does not seem to break symmetries so far proposed in literatures because of the geometrical frustration, while the full understanding of the spin liquid phase remains a challenge. 
Most of later numerical~\cite{Kyung} and theoretical~\cite{Lee} studies have also been performed for a simplified single-band 2D Hubbard model based on an empirical estimate of parameters by following extended H\"uckel calculations.~\cite{Mori,Saito}  
We certainly need a more realistic and {\it ab initio} description of $\kappa$-ET compounds to establish the existence of such truly as-yet-unestablished states and to get insights into possible fundamentally new ideas.  

Another seminal finding achieved in this family is the unconventional character of the Mott transition found for $X$=Cu[N(CN)$_2$]Cl (abbreviated as $\kappa$-Cl) under pressure.~\cite{Kagawa}   
The novel universality class observed by the resistivity at this Mott transition is in good agreement with the marginal quantum criticality at the meeting point of the symmetry breaking and topological change.~\cite{Imada1,Imada2,Misawa1,Misawa2}
Its significance to physics and theoretical concept of the quantum criticality calls further experimental test and critical examinations based on the comparison with the realistic and first-principles grounds.  
 
In spite of these innovative ideas and findings, the {\it ab initio} studies are so far few~\cite{DFT_ET} and most of the studies were performed using empirical models inferred from the H\"uckel studies. In addition, the strong electron correlation in the organic conductors hardly justifies its naive applications of the 
standard density functional theory (DFT).   

To overcome the limitation and deficiency of the conventional DFT approach, a hybrid first-principles framework, which we call Multi-scale {\it Ab initio} scheme for Correlated Electrons (MACE) has been developed and 
applied to a number of strongly correlated materials.~\cite{Imada-Miyake}
Our general framework consists of (1) {\it ab initio} 
calculations of the global electronic band structure
either by the density functional theory or by other many-body theory such as the GW method and (2) a subsequent downfolding procedure by elimination of degrees of freedom far away from the Fermi level, which generates low-energy effective models.~\cite{Aryasetiawan,Solovyev,Imai,Otsuka} It is followed by (3) the procedure to solve the low-energy models by more reliable low-energy solvers such as  PIRG, dynamical mean-field theory(DMFT),~\cite{georges} and the many-variable variational Monte Carlo (mVMC) methods.~\cite{Tahara}
The accuracy of this three-stage scheme has critically been tested widely against various cases of exciton excitations in semiconductors,~\cite{NakamuraGaAs} phase diagrams with competing orders in transition metal compounds,~\cite{Imai,Otsuka} and iron superconductors.~\cite{Misawa,Nakamura,Miyake} A specific combination of the local density approximation (LDA) for the first step and DMFT for the final step has also been widely applied\cite{LDA+DMFT}. They have been favorably compared with available experimental results.

However, the accuracy is not clear for more complex compounds and for more strongly correlated systems. Indeed applications to the organic conductors remain a grand challenge. The {\it ab initio} low-energy models have recently been derived for the $\kappa$-ET compounds along the line of the above first and second parts of the three-stage scheme in  the present terminology of MACE.~\cite{NakamuraET} In fact, they have shown that the ratio of the typical correlation strength (local screened Coulomb interaction), $U$,  to the typical electron transfer $t$ is as large as 10 for the downfolded single band models.  Furthermore, we have found that the ratio of the typical off-site Coulomb interaction $V$ to $U$ is as large as $1/3$. Elucidating electronic structures under such a strong correlation is a challenge at the forefront of research to develop efficient and accurate numerical algorithms.  

The purpose of this study is to make a further step to the third procedure following the spirit of MACE; we solve the {\it ab initio} effective Hamiltonian of real $\kappa$-ET compounds by using an accurate solver based on a recently developed and improved mVMC method with many variational parameters~\cite{Tahara} to see whether the present general framework combined with the mVMC solver offers an accurate framework for the complex and strongly correlated organic conductors.  
For this purpose,
we study $\kappa$-NCS, as a typical compound close to the Mott transition.
 
This compound is in fact barely metallic with an enhancement of the antiferromagnetic correlations revealed by the nuclear relaxation rate $T_1$ and located close to the antiferromagnetic Mott insulating phase.~\cite{Kanoda} In fact, above 90 K,~\cite{Kawamoto} the resistivity shows insulating-like increase with decreasing temperatures and has an inflection point around $T$=55 K (ref.~\citen{Urayama}) coinciding with the peak of $1/(T_1T)$ with a crossover to a metallic behavior below it.  It eventually becomes superconducting below around 10 K.  

In this study, we assume normal states (not the superconductor) for the candidate of the metal, while leave it arbitrary for the insulator to study competitions between metals, antiferromagnetic or charge ordered insulators as well as the Mott insulator without symmetry breakings. We examine the phase boundary between the metal and the Mott insulator in a parameter space by taking the relative electron correlation amplitude as a parameter beyond the {\it ab initio} value. In practice, we draw a phase diagram as a function of a parameter $\lambda$ that monitors the relative strength of the effective interaction to the electron transfer by uniformly scaling the interaction strength, where the {\it ab initio} value is given by $\lambda=1$. 
It gives us an idea about the relative location of the real material to this phase boundary and hence the relevance of the Mott physics. As we will show in \S 3, off-site Coulomb and exchange interactions largely stabilize a paramagnetic metal in the region of strong correlation ($U/t\geq 6$). 

The present work is, to our knowledge, the first {\it ab initio} attempt to estimate the Mott transition and its neighboring phases in organic conductors containing a large number of atoms beyond 100 with four complex ET molecules in a unit cell. Our results indicate that $\kappa$-NCS is indeed near the Mott transition within the accuracy of 20\%. It also shows that antiferromagnetic insulating state exists even in the metallic phase near the Mott transition, as a metastable excited phase, whose energy is typically about 1~meV $\sim$ 10 K higher than that of the paramagnetic phase. This is also consistent with the above experimental results of the crossover between high-temperature insulating and low-temperature paramagnetic metallic phases. Although the results show overall agreement with the experiments, a closer look of the ground state obtained with the realistic {\it ab initio} parameter (i.e., $\lambda=1$) becomes an antiferromagnetic Mott insulator in contrast to the metallic phase 
in the experimental indications. We then discuss possible origins of this discrepancy.

This paper is organized as follows: We briefly summarize in \S 2 the method of our calculation with {\it ab initio} downfolding of the low-energy model for $\kappa$-NCS and the basic framework of the mVMC method we employed. In \S 3 calculated results are presented. The summary and discussions are given in \S 4. 

\section{Method}

\subsection{Derivation of low-energy effective model} 
Here, we describe a derivation of low-energy effective models for the present system. The scheme is based on first principles calculations and an application of the first two stages of the three-stage scheme. The basis of the Hamiltonian is the Wannier function associated with antibonding states of the highest occupied molecular orbitals (HOMOs) of two ET molecules that form a dimer. In the present paper, we restrict our consideration within the single-band models, where we derive the model containing only the degrees of freedom for the antibonding band while the bonding band is traced out in the downfolding. Possible dynamical effects of the bonding degrees of freedom remain as future issues. The explicit form of this Hamiltonian is given in the form of the two-dimensional (2D) single-band extended Hubbard model as 
\begin{eqnarray}
	\mathcal{H}\!&=&\! -\sum_{\sigma} \sum_{i\neq j} t_{ij} a_{i \sigma}^{\dagger} a_{j \sigma}+\frac{1}{2} \sum_{\sigma \rho} \sum_{i,j} V_{ij} a_{i \sigma}^{\dagger} 
  a_{j \rho}^{\dagger} a_{j \rho} a_{i \sigma}  \nonumber \\ 
	\!\!\!&+&\!\!\!\frac{1}{2} \sum_{\sigma \rho} \sum_{i\neq j} J_{ij} \left(a_{i \sigma}^{\dagger} a_{j \rho}^{\dagger} a_{i \rho} a_{j \sigma} + a_{i \sigma}^{\dagger} a_{i \rho}^{\dagger} a_{j \rho} a_{j \sigma}\right),
\label{H_Hub}
\end{eqnarray}
where $a_{i \sigma}^{\dagger}$ ($a_{i \sigma}$) is a creation (annihilation) operator of an electron with spin $\sigma$ in the Wannier orbital localized at the $i$th BEDT-TTF dimer. The $t_{ij}$\ parameters are given by 
\begin{eqnarray}
	t_{ij} = \langle \phi_{i} | \mathcal{H}_\mathrm{KS} |  \phi_{j} \rangle \label{t_ij} 
\end{eqnarray}
with $| \phi_{i} \rangle =a_{i}^{\dagger}|0\rangle$ and $\mathcal{H}_\mathrm{KS}$ being the Kohn-Sham Hamiltonian representing an effective one-body potential.
The $V_{ij}$ and $J_{ij}$ parameters are screened Coulomb and exchange integrals in the Wannier-orbital basis, respectively, expressed as 
\begin{eqnarray}
V_{ij}
\!&=&\!\langle \phi_{i} \phi_{j} | W | \phi_{i} \phi_{j} \rangle \nonumber \\ 
\!&=&\!\int \int d{\bf r} d{\bf r}' \phi_{i}^{*}({\bf r}) \phi_{i}({\bf r}) 
W({\bf r},{\bf r}') \phi_{j}^{*}({\bf r}') \phi_{j}({\bf r}') 
\label{V_ij}
\end{eqnarray} 
and
\begin{eqnarray}
J_{ij} 
\!&=&\!\langle \phi_{i} \phi_{j} | W | \phi_{j} \phi_{i} \rangle \nonumber \\ 
\!&=&\!\int \int d{\bf r} d{\bf r}' \phi_{i}^{*}({\bf r}) \phi_{j}({\bf r}) 
W({\bf r},{\bf r}') \phi_{j}^{*}({\bf r}') \phi_{i}({\bf r}') 
\label{J_ij}
\end{eqnarray}
with $W({\bf r},{\bf r}')$ being a 2D screened Coulomb interaction in the low-frequency limit.
Note that the Hamiltonian given in eq.~(\ref{H_Hub}) can be rewritten as
\begin{eqnarray}
	\mathcal{H}\!&=&\! -\sum_{\sigma} \sum_{i\neq j} t_{ij} a_{i \sigma}^{\dagger} a_{j \sigma}  + \sum_{i} U n_{i\uparrow}n_{i\downarrow} \nonumber \\
	\!\!\!&+&\!\!\!\frac{1}{2}\sum_{i\ne j} \left(V_{ij}-\frac{1}{2}J_{ij}\right) n_i n_j\nonumber \nonumber \\
	\!\!\!&-&\!\!\! \sum_{i\neq j} J_{ij} \left(S_i\cdot S_j - a_{i\uparrow}^\dagger a_{j\uparrow} a_{i\downarrow}^\dagger a_{j\downarrow}\right),
\label{H_Hub_transform}
\end{eqnarray}
where the onsite Hubbard parameter $U$ is given by $V_{ii}$, $n_{i\sigma} = a_{i\sigma}^\dagger a_{i\sigma}$ and $n_i = n_{i\uparrow}+n_{i\downarrow}$.
The spin operator $S_i$ is defined as $S_i = (S_i^x, S_i^y, S_i^z)$, where $S_i^x/\mathrm{i}S_i^y = (a_{i\uparrow}^\dagger a_{i\downarrow} \pm a_{i\downarrow}^\dagger a_{i\uparrow})/2$ and $S_i^z = (n_{i\uparrow}-n_{i\downarrow})/2$.

The derivation of $W$ for the purely 2D system follows ref.~\citen{Nakamura2D}, where a new framework of the constrained random-phase approximation (cRPA) was developed for the purpose to derive effective interactions of models defined in lower spatial dimensions. This new scheme is suitable for quasi-low-dimensional materials such as the present system. The cRPA method is originally formulated in the RPA framework with the constraint for the band degree of freedom 
to eliminate only the degrees of freedom far from the Fermi level in energy.
This is called the {\it band downfolding}. In the proposed scheme of the supplementary downfolding~\cite{Nakamura2D}, however, the concept of the constraint is additionally relaxed to include the screening by the polarization in the other layers/chains even within the target bands. This is formulated in the {\it real space representation} and eliminates
the degrees of freedom away from the target layer/chain, which results in the low-dimensional model for the target layer/chain. We call it the {\it dimensional downfolding}. 

Practically, the band+dimensional downfolding is performed in two steps: We first perform the band downfolding to derive the 3D model for small number of bands near the Fermi level.~\cite{NakamuraET} This is followed by the dimensional downfolding in the second step.~\cite{Nakamura2D} With this idea, we can naturally derive the low-energy model in any dimensions. In the present case, we use it for the derivation of a 2D model for $\kappa$-SCN.

\subsection{Multi-variable variational Monte Carlo method}
To investigate ground-state properties of the low-energy 2D model, we employ a multi-variable variational Monte Carlo method (mVMC) combined with quantum-number projection and multi-variable optimization~\cite{Tahara, Tahara2}.
The variational wave function $|\psi\rangle$ is defined as
\begin{eqnarray}
 |\psi \rangle &=& \mathcal{P}_\mathrm{J} \mathcal{P}^\mathrm{ex.}_\mathrm{d-h} \mathcal{P}_\mathrm{G} \mathcal{L}^{S=0} |\phi_\mathrm{pair}\rangle,
\end{eqnarray}
where $\mathcal{P}_\mathrm{G}$, $\mathcal{P}^\mathrm{ex.}_\mathrm{d-h}$, and $\mathcal{P}_\mathrm{J}$ are the Gutzwiller factor~\cite{Gutzwiller}, the doublon-holon correlation factor~\cite{Kaplan, YokoyamaShiba3}, and the Jastrow factor~\cite{Jastrow}, respectively.
These factors are defined as
\begin{eqnarray}
  \sP_{\text{G}} &=& \exp \biggl[ -g\sum_{i} n_{i\uparrow} n_{i\downarrow} \biggr], \\
  \sP_{\text{d-h}}^{\text{ex.}} &=& \exp\biggl[ - \sum_{m=0}^{2}\sum_{\ell=1,2} \ga_{(m)}^{(\ell)} \sum_{i} \xi_{i(m)}^{(\ell)} \biggr], \\
  \sP_{\text{J}} &=& \exp \biggl[ - \frac{1}{2} \sum_{i\neq j} v_{ij} \bigl(n_{i\uparrow}+n_{i\downarrow}\bigr) \bigl(n_{j\uparrow}+n_{j\downarrow}\bigr) \biggr] ,
\end{eqnarray}
where $g$, $\ga_{(m)}^{(\ell)}$, and $v_{ij}$ are variational parameters. 
Here, $\xi_{i(m)}^{(\ell)}$ is a many-body operator which is diagonal in the real-space representations. 
When a doublon (holon) exists at the $i$-th site and $m$ holons (doublons) surround at the $\ell$-th nearest neighbor, $\xi_{i(m)}^{(\ell)}$ gives $1$. Otherwise, $\xi_{i(m)}^{(\ell)}$ gives $0$. 
In the present study, we take $v_{ij} = v(r_{ij}) = v(-r_{ij})$, where $r_{ij} = \vr_i-\vr_j$ is the relative displacement between the sites $i$ and $j$.
The spin quantum-number projection operator $\mathcal{L}^{S=0}$ restores the $SU(2)$ spin-rotational symmetry with the total spin $S=0$~\cite{ManyBody, PIRGMizusaki}.

For the one-body part $|\phi_\mathrm{pair}\rangle$, we employ a generalized pairing wave function defined as
\begin{eqnarray}
	|\phi_\mathrm{pair}\rangle = \left( \sum_{i,j=1}^{\Ns} f_{ij} c_{i\uparrow}^\dagger c_{j\downarrow}^\dagger \right)^{N/2}|0\rangle,\label{eq:one-body-wf}
\end{eqnarray}
where $f_{ij}$ and $\Ns$ are variational parameters and the total number of the sites, respectively.
The number of electrons is denoted by $N$.
Throughout this paper, we consider the half-filled case $N=N_\mathrm{s}$.
The paring wave function given in eq.~(\ref{eq:one-body-wf}) can flexibly describe 
paramagnetic metals/insulators, antiferromagnetic insulators/metals, charge-ordered insulators/metals, superconducting phases as well as phases with strong quantum fluctuations showing developed spin and/or charge correlations decaying with any power laws as functions of distance.
In fact, the mVMC method based on the variational function in eq.~(6) can describe paramagnetic metals with developed spin correlations in hole-doped Hubbard models on square lattices.~\cite{Tahara}

\begin{figure}
 \centering
 \resizebox{0.3\textwidth}{!}{\includegraphics{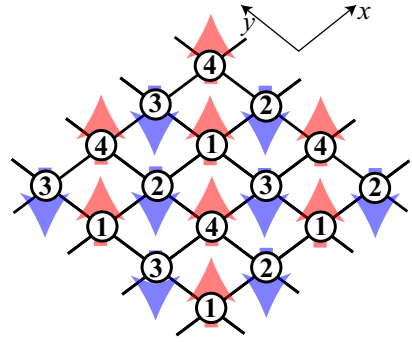}}
 \caption{(Color online) Schematic illustration of a $4\times 4$ super cell.
 The labels $1$--$4$ denote the sublattice indices for the $2 \times 2$ sublattice structure for $f_{ij}$.
 The arrows represent the staggered magnetization in the AFI (see Fig.~\ref{fig:lambda-2D}).}
 \label{fig:sblattice-2x2}
\end{figure}

\begin{figure}
 \centering
 \resizebox{0.4\textwidth}{!}{\includegraphics{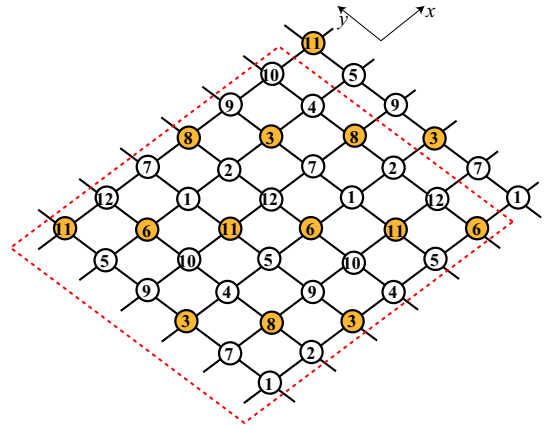}}
 \caption{(Color online) (a) Schematic of the three-fold charge-ordered (CO) phase. The colored (shaded) and white sites denote charge-rich and charge-poor sites, respectively. The labels denote the sublattice indices for the $6\times 2$ sublattice structure for $f_{ij}$.}
 \label{fig:sblattice-6x2}
\end{figure}

In principle, it is better to allow variational wavefunctions as much as flexible  without imposing any constraint on $f_{ij}$. However, if we do not impose the constraint, the possible variational parameters $f_{ij}$ increase in proportion to $N_s^2$, which increases the computation time for the optimization of the variational parameters enormously for large system sizes. To reduce the computational cost, we restrict $f_{ij}$ to have a $\ell_{sx} \times \ell_{sy}$ sublattice structure as $f_{ij} = f_{\sigma({j})}(\vr_i - \vr_j)$, where $\sigma(j)$ is a sublattice index at site $j$. In the study of the Mott transition in \S \ref{Mott transition}, we take $\ell_{sx}=\ell_{sy}=2$ (see Fig.~\ref{fig:sblattice-2x2}).
This assumption is based on the observation of the staggered magnetization with ordering vector of $(\pi,\pi)$ in the frustrated Hubbard model at $t^\prime/t\le0.6$~\cite{Mizusaki,Morita}, and the spin-$\frac{1}{2}$ quantum Heisenberg antiferromagnets corresponding to $U\rightarrow +\infty$ for $J_2/J_1<(0.6)^{2}$.~\cite{Dagotto89} 
These appear to be relevant to the present parameter region as we will see in the comparison with the exact diagonalization as well.
For the study of the charge order in \S \ref{Charge-ordered insulator} (as we describe in detail there), we take more general $\ell_{sx}=6$ and $\ell_{sy}=2$ to allow the three-sublattice as well as two-sublattice structures (See Fig.~\ref{fig:sblattice-6x2}).

In the following calculations, we take $\Ns=L\times L$ sites with periodic boundary conditions.
All the variational parameters are simultaneously optimized by using the stochastic reconfiguration method~\cite{Sorella-SR}.

\section{Results}
In this section, we show our computed results including {\it ab initio} band calculations, derived transfer and interaction parameters in the low-energy 2D model for $\kappa$-NCS, and analyses for this model with the mVMC method. 
Our calculation has been performed by using the experimental lattice structure of $\kappa$-NCS taken from the neutron diffraction data at 15 K by Schultz {\it et al.}\cite{Schultz}

\subsection{Low-energy model}
The present {\it ab initio} calculations were performed with an electronic-structure code based on plane-wave basis set, {\it Tokyo Ab initio Program Package}.~\cite{Ref_TAPP} Density-functional calculations with the generalized gradient approximation (GGA) using the Perdew-Burke-Ernzerhof parameterization~\cite{Ref_PBE96} were performed with the Troullier-Martins norm-conserving pseudopotentials~\cite{Ref_PP1} in the Kleinman-Bylander representation.\cite{Ref_PP2} The cutoff energies in wavefunctions and charge densities were set to 36 Ry and 144 Ry, respectively. We employed a 5$\times$5$\times$5 $k$-point sampling for the Brillouin-zone integral.
The electronic structure with the cutoff energy of 36 Ry was compared with the higher cutoff of 49 Ry. We confirmed that the 36-Ry band dispersion is almost identical to the 49-Ry one. We also confirmed that the wavefunctions of the BEDT-TTF molecules in the low-energy, being essential in the polarization calculation, are well converged for the present cutoff.
The construction of the maximally-localized Wannier functions follows ref.~\citen{Marzari}. The polarization function was expanded in plane waves with an energy cutoff of 5 Ry and the total number of bands considered in the polarization calculation was set to 750, where the numbers of occupied, partially-occupied, and unoccupied bands are 232, 2, and 416, respectively.
This condition corresponds to considering excitations up to $\sim$21 eV above the Fermi level.
The Brillouin-zone integral on wavevectors was evaluated by the generalized tetrahedron method.\cite{Fujiwara} The additional terms in the long-wavelength polarization function due to nonlocal terms in the pseudopotentials were explicitly considered following ref.~\citen{Louie}. In the evaluation of the Wannier matrix elements, $V_{ij}$ and $J_{ij}$, the singularity in the Coulomb interaction at the long-wave-length limit was treated in the manner described in ref.~\citen{Louie}. We confirmed that these conditions give well converged results.

\begin{figure*}
\centering
\resizebox{0.7\textwidth}{!}{\includegraphics{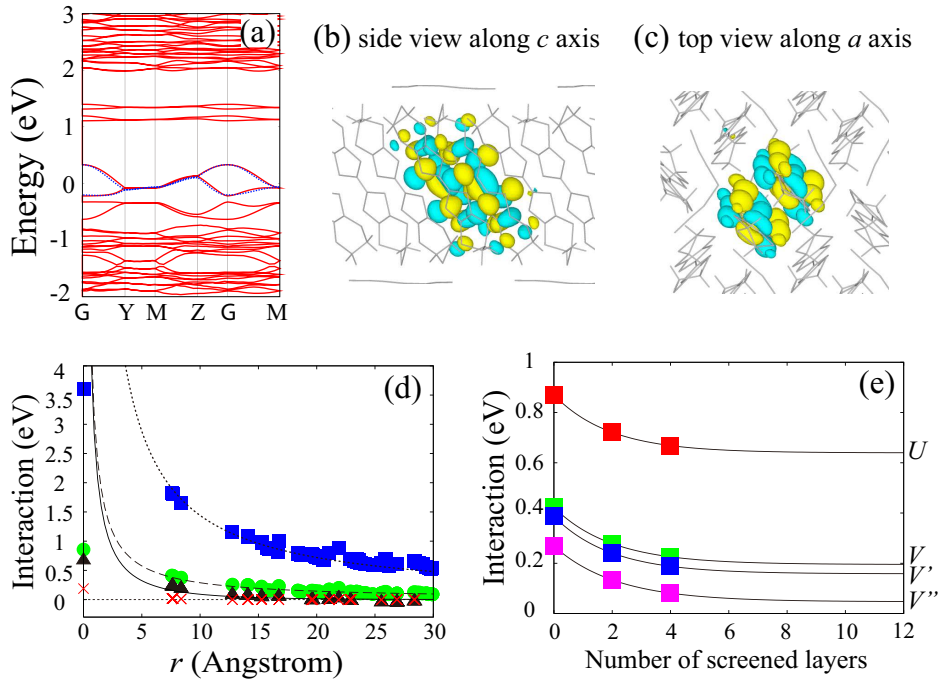}}
\caption{(Color online) (a) Calculated GGA band structures (red line) of $\kappa$-(BEDT-TTF)$_2$Cu(NCS)$_2$. The crystal structure contains alternating layers (parallel to the $bc$ plane) of BEDT-TTF donor molecules and polymeric Cu(NCS)$_{2}^{-}$ anions. Band dispersions are plotted along the high-symmetry points in the $bc$ plane, where $\Gamma$ = (0, 0, 0), Y = (0, $b^{*}$/2, 0), Z = (0, 0, $c^{*}$/2), and M = (0, $b^{*}$/2, $c^{*}$/2).
Note that the $a$ axis is interlayer axis.
The zero of energy is the Fermi level. The (blue) dotted dispersions are obtained by the three transfer parameters listed in Fig.~\ref{transfer}(a). The panels (b) and (c) display isosurface contours of maximally localized Wannier functions for the target band, where (b) and (c) show the side and top views, respectively. The amplitudes of the contour surface are 0.02 [light grey (yellow)] and $-$0.02 [dark grey (blue)] in the atomic unit.
(d) Calculated screened Coulomb interactions of $\kappa$-(BEDT-TTF)$_2$Cu(NCS)$_2$ as a function of the distance between the centers of maximally localized Wannier orbitals. The (blue) squares, (green) circles, (black) triangles, and (red) crosses represent the bare, 3D-cRPA, 2D-cRPA, and full-RPA interactions, respectively. The dotted, dashed, and solid curves denote $s/r$, $s/\epsilon r$, and $s \exp(-r/\sigma)/\epsilon r$, respectively, where a decay constant $\epsilon = 5.0$ was determined by the fitting to the 3D-cRPA data. Also, $s$ is a unit parameter of 14.40 eV$\cdot$\AA and $\sigma$ is the characteristic screening length of the 2D-cRPA interaction, corresponding to the interlayer distance of 16.4 \AA.
(e) Convergence of interaction parameters $V_{ij}$ up to the third neighbors as functions of number of screening layers $n$ contained in a unit cell. $V_{ij}$ are extrapolated to $n\rightarrow \infty$. Solid curves are the fitted exponential functions.
Here, $V,V'$, and $V''$ represent the first, second and third neighbor interactions, respectively.}
\label{Interaction}
\end{figure*} 
\begin{figure*}
 \centering
 \resizebox{0.8\textwidth}{!}{\includegraphics{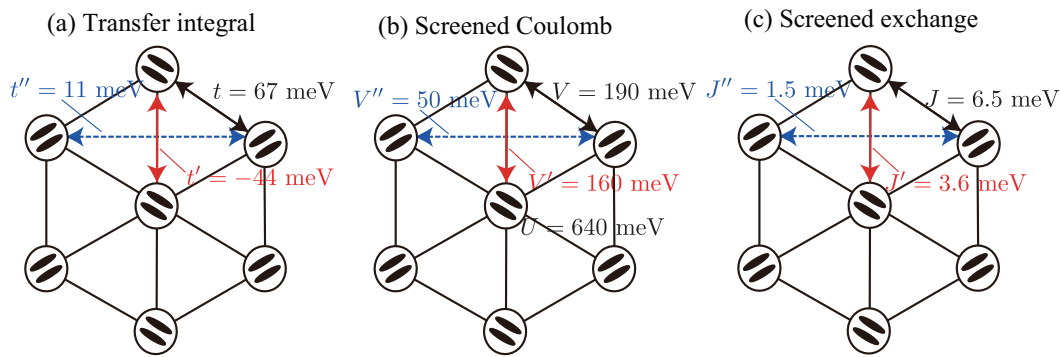}}
 \caption{(Color online) Schematic illustration of 2D BEDT-TTF layer, where a BEDT-TTF molecule is described as an ellipsoid and its dimer is written as a circle. In the dimer limit, the system forms anisotropic triangular lattice. The derived parameters for the 2D extended Hubbard model in eq.~(\ref{H_Hub}) are shown for transfer integrals (a) and screened Coulomb (b) and screened exchange (c) interactions up to the third neighbors. }
 \label{transfer}
\end{figure*} 
Figure~\ref{Interaction}(a) shows our calculated GGA band structure of $\kappa$-NCS. (Blue) dotted lines are a tight-binding band with three transfers up to the third nearest neighbors, listed in Fig.~\ref{transfer}(a), derived with matrix elements of the Kohn-Sham Hamiltonians in the maximally-localized Wannier-orbital basis. The resulting bandwidth of the target band for the effective model is 0.56 eV. We note that interlayer transfers are considerably small as $\sim$ 0.1 meV compared to intralayer transfers $\sim$ 65 meV and thus the present system is regarded as a typical quasi-2D system. 
Figures~\ref{Interaction}(b) and (c) are visualization of our calculated Wannier function in the side and top views, respectively.
From the figures, we see that the Wannier orbital is the anti-bonding state of HOMOs of two ET molecules and confined in the layers.

Figure~\ref{Interaction}(d) shows the calculated cRPA interactions plotted as a function of distance between centers of the Wannier orbitals. (Green) filled circles represent the conventional cRPA result based on ``band downfolding'' scheme. This cRPA interaction exhibits a power-law decay (dashed lines)
\begin{eqnarray} 
f(r) = \frac{s}{\epsilon r} 
\label{long-range}
\end{eqnarray} 
at long distances with the dielectric constant $\epsilon$ = 5.0 and the unit parameter $s$ = 14.40 eV$\cdot$\AA.

On the other hand, (black) triangles in Fig.~\ref{Interaction}(d) describe the result obtained from the present ``band + dimensional downfolding'' scheme. 
After this dimensional downfolding, the effective interaction becomes qualitatively different from the long-ranged form and is reduced to a short-ranged interaction. In fact, it fits well with the Yukawa type form (solid lines) as
\begin{eqnarray} 
g(r) = \frac{s \exp(-r/\sigma)}{\epsilon r},
\label{Yukawa}
\end{eqnarray}
where $\sigma=16.4\AA$ is the interlayer distance between the ET layers. This qualitative change into the short-ranged interaction comes from the screening by the gapless polarization channel of other metallic layers, which enters in the dimensional downfolding process. To explicitly distinguish, hereafter, we refer to the former yielding eq.~(\ref{long-range}) as 3D-cRPA and to the latter with eq.~(\ref{Yukawa}) as 2D-cRPA. For comparison, the figure includes the bare ((blue) squares) and full-RPA ((red) crosses) results as well.
The dotted line is the bare Coulomb interaction decaying as $s/r$.

On the basis of this exponential dependence in the 2D-cRPA, it is justified to employ the Hubbard-type model with only the short-ranged interaction. Practically, $V_{ij}$ is considered up to the third nearest neighbors in this paper, because $V_{ij}$ is negligible beyond this range. We also note that the interlayer screening affects even the onsite interaction, reducing $U$=0.86 eV for 3D-cRPA to $U$=0.64 eV for 2D-cRPA by $\sim$ 25 \%.
For reference, we note that the full-RPA $U$ is 0.19 eV and thus the intralayer screening reduces the effective Hubbard $U$ further by 70 \%.

In the calculation of the interlayer screening for the dimensional downfolding, we consider stacked supercells, each of which contains one target layer/chain. Electrons on this target layer/chain are screened by those on other layers. The supercell is employed just for the technical reason of the calculation and the supercell size should be extrapolated to the infinity afterwards.
Note that the size of the superlattice $N_\mathrm{L}$ is nothing but the number of the sampling-$k$ points along the $a^{*}$ axis perpendicular to the layer.
In the present case, we consider the case where the total system consists of the supercell containing up to five stacking layers (one target layer and up to four screening layers), where we sample five $k$-points along the interlayer $a^{*}$ axis for the case of the maximum supercell size.
In Fig.~\ref{Interaction}(e), we extrapolate $V_{ij}$ to the thermodynamic limit of $N_\mathrm{L}=+\infty$ with an exponential function.
The obtained $V_{ij}$ in the thermodynamic limit are summarized in Fig.~\ref{transfer}(b).
The estimated $U/t$ is as large as 9.6, apparently supporting that $\kappa$-NCS belongs to a material with strongly correlated electrons.

Screened (direct) exchange interactions $J_{ij}$ are also calculated with 2D-cRPA. The results are shown in Fig.~\ref{transfer}(c).
The exchange interactions are hardly screened~\cite{Miyake,Nakamura} and have no significant system-size dependence.
One might think that $J_{ij}$ are negligibly small compared to the Coulomb interactions $V_{ij}$.
However, we will see that the exchange interactions have a discernible effect on the critical interaction ratio of the metal-insulator transition as demonstrated in Sec.~\ref{sec:MIT}.
For the nearest neighbor pairs in the strong coupling limit, the kinetic exchange is estimated by $-2|t_{ij}|^2/(U-V_{ij})$ as $-$20.0 meV.
This value is comparable to the nearest-neighbor direct exchange interaction $J_{ij}=6.5~\mathrm{meV}$.
Indeed, $J_{ij}$ and $V_{ij}$, which are absent in the simple Hubbard model, play a substantial role in quantitatively determining electronic properties of $\kappa$-NCS as demonstrated in the next section.

\subsection{Ground-state properties of the low-energy model}\label{sec:MIT}
We now present the ground-state properties of our derived 2D low-energy model obtained by using the mVMC method after extrapolating finite-size results to the thermodynamic limit.

\subsubsection{Ground-state phase diagram}
\begin{figure}
 \centering
\resizebox{0.4\textwidth}{!}{\includegraphics{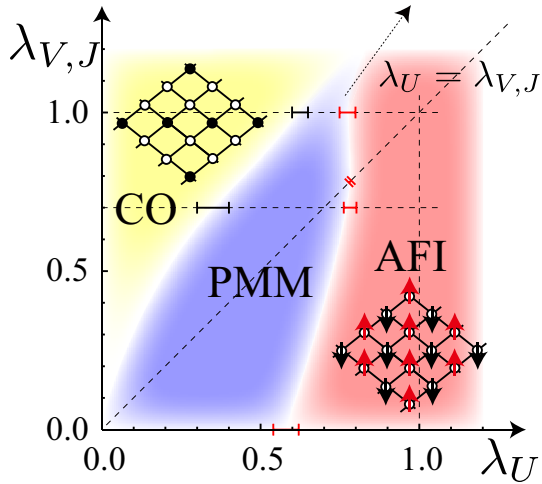}}
\caption{(Color online) Calculated $\lU$-$\lV$ ground-state phase diagram of the model given in eq.~(\ref{H_Hub}). PMM, AFI and CO denote a paramagnetic metal, an antiferromagnetic insulator and a charge-ordered phase, respectively.
The transition points on the diagonal line and the horizontal line of $\lV=0$ are those in the thermodynamic limit.
 The transition points on the horizontal lines of $\lV=1$ and $0.7$ are determined by the data for $L=12$, which are expected to be close to the thermodynamic limit.
 The arrow with dotted line between CO and AFI denotes the transition line between uniform and three-fold charge-ordered phases analytically estimated in the limit of $\lU, \lV\rightarrow +\infty$.}
 \label{fig:lambda-2D}
\end{figure}

For a comprehensive understanding of low-energy electronic structures of the {\it ab initio} model, we work out the ground-state phase diagram in the parameter space of $\lU$ and $\lV$, where $U$ is scaled from the realistic value by the factor $\lU$, while $V_{ij}$ and $J_{ij}$ are scaled by the factor $\lV$. 
The obtained phase diagram is illustrated in Fig.~\ref{fig:lambda-2D}.
In the weakly-correlated region (namely in the region of $\lU\ll 1$ and $\lV\ll 1$), the ground state is a paramagnetic metal (PMM). 
With increasing the interactions along the diagonal line of $\lambda$=$\lU$=$\lV$, the system undergoes, at $\lambda$$\simeq$0.782$\pm$0.005, a Mott transition into an antiferromagnetic insulator (AFI) with the ordering vector at $(\pi,\pi)$. Thus, the present three-stage approach reasonably reproduces the experimental fact that $\kappa$-NCS is on the verge of the metal-insulator transition.~\cite{Kagawa}

The phase diagram tells us that the off-site interactions $V_{ij}$ and $J_{ij}$ largely stabilize the paramagnetic metal; for the case of $\lV=0$, the system turns into the Mott insulator at $\lU\simeq0.58\pm0.04$, which is 20\% smaller than that for the diagonal line. 
On the other hand, a charge-ordered (CO) phase emerges for $\lU\lesssim \lV$.
In the CO phase, the system exhibits a three-fold (\textit{rich-poor-poor}) charge order with ordering vector of $(2\pi/3, 2\pi/3)$, which reduces the nearest-neighbor Coulomb repulsion energy (see an inset of Fig.~\ref{fig:lambda-2D}). The nature of the CO phase will be discussed in \S \ref{Charge-ordered insulator}.

\subsubsection{Mott transition} \label{Mott transition}
In this subsection, we discuss the Mott transition between the paramagnetic metal and the antiferromagnetic insulator.
Let us start with analyses on the diagonal line of $\lU=\lV$.
To identify quantum phase transitions, we calculate the doublon density
\begin{eqnarray}
	D &\equiv& \frac{1}{\Ns}\sum_i \langle n_{i\uparrow}n_{i\downarrow}\rangle,
\end{eqnarray}
the momentum distribution $n(\bm{k})$
\begin{eqnarray}
	n(\bm{k}) &\equiv& \frac{1}{2\Ns} \sum_{ij\sigma} \langle c^\dagger_{i\sigma}c_{j\sigma} \rangle e^{-{\rm i}\bm{k}\cdot (\bm{r}_i-\bm{r}_j)},
\end{eqnarray}
and the spin structure factor
\begin{eqnarray}
	S(\bm{q}) &\equiv& \frac{1}{3\Ns} \sum_{ij} \langle \vec{S}_i \cdot \vec{S}_j \rangle e^{-{\rm i}\bm{q}\cdot (\bm{r}_i-\bm{r}_j)}.
\end{eqnarray}

\begin{figure*}
 \centering
 \resizebox{0.6\textwidth}{!}{\includegraphics{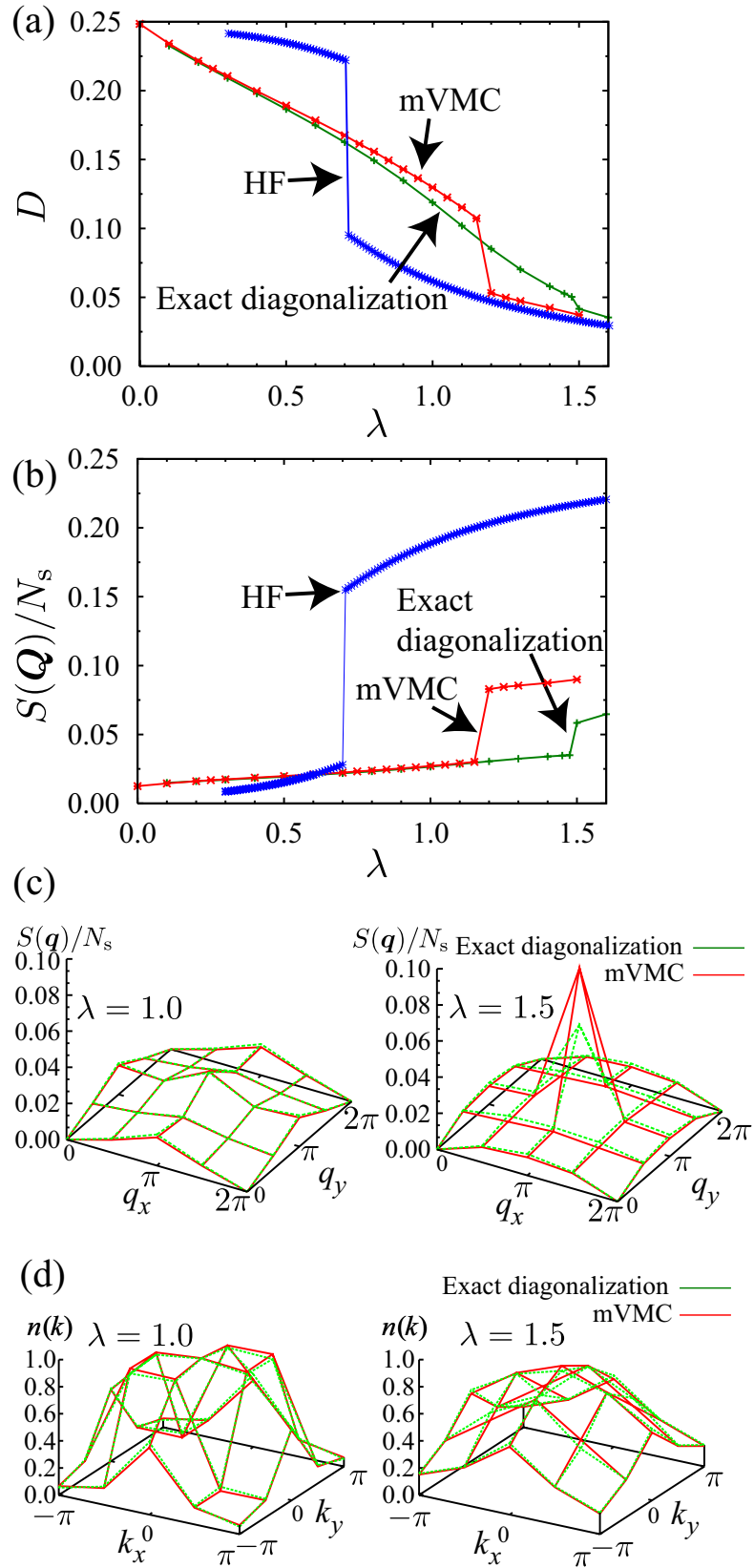}}
 \caption{(Color online) 
Calculated doublon density $D$ and spin structure factor $S(\bm{Q})$ for $L=4$. Here $Q$ denotes $(\pi, \pi)$. For details, see text. (c)  Spin structure factor $S(\bm{q})$ and (d) momentum distribution $n(\bm{k})$ at $\lambda=1.0$ and 1.5.}
 \label{fig:phys-L4}
\end{figure*}

In Fig.~\ref{fig:phys-L4}, we compare calculated results obtained by the unrestricted Hartree-Fock (UHF), and mVMC calculations with those by exact-diagonalization methods for $L=4$,
to roughly get insight into accuracies of the mVMC and UHF methods.
Hereafter we measure system sizes in units of the unit cell consisting of a square constructed from the nearest-neighbor bonds, in which the linear dimension is denoted by $L$.
In all of the three methods, the results show qualitatively similar behaviors, where by turning on $\lambda$, $D$ decreases continuously from 0.25 and suddenly exhibits a discontinuous decrease at $\lambda_\mathrm{c}$ [panel (a)].
At the same time, $S(\bm{Q})/\Ns$ shows a jump at $\lambda=\lc$ [(b)].
For $\lambda>\lc$, a clear single peak emerges at $\bm{Q}=(\pi,\pi)$ in $S(\bm{q})/\Ns$ as shown in Fig.~\ref{fig:phys-L4}(c).
We found no other anomaly in $S(\bm{q})$ and $D$. 
These numerical results suggest that the system is under the influence of the electron correlations and undergoes a first-order transition from the paramagnetic metal to the antiferromagnetic insulator at $\lc$.
For the exact diagonalization, we obtain $\lc\sim1.49$ for $L=4$.
On the other hand, the mVMC result indicates $\lc=1.15$--$1.2$, which is about 20\% smaller than the exact value.
Despite the underestimate of the critical interaction ratio, the mVMC well reproduces $S(\bm{Q})$, $D$, and $n(\bm{k})$ in the AFI and PMM phases [see also Fig.\ref{fig:phys-L4}(d)].
It should be mentioned that a UHF calculation gives $\lc=0.6$--0.65, which is over 50\% smaller than the exact value.
To summarize, the present small cluster result implies that 
the mVMC solver tends to underestimate $\lc$ by as much as 20\% compared to the exact-diagonalization solver.
The comparison between the mVMC and PIRG results for the Hubbard model on the square lattice with the next-nearest-neighbor transfers has shown the underestimate of $\lc$ for the metal-insulator transition by the mVMC as much as 10 \% in the thermodynamic limit.\cite{Tahara2} 
We further discuss the implication of this discrepancy in \S \ref{Summary}.

\begin{figure}
 \centering
\resizebox{0.425\textwidth}{!}{\includegraphics{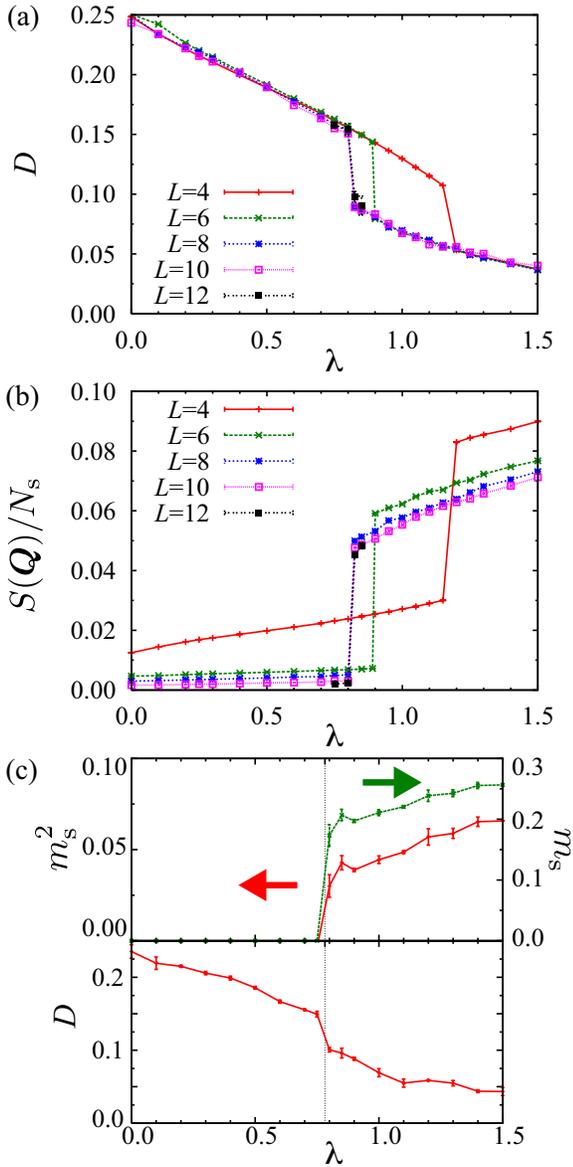}}
\caption{(Color online) System-size dependences of (a) doublon density $D$, and (b) spin structure factor $S(\bm{Q})$. Here $Q$ denotes $(\pi, \pi)$. (c) $D$, $m_\mathrm{s}^2$, and $m_\mathrm{s}$ extrapolated to the thermodynamic limit. All the data are for $\lU=\lV$ which corresponds to the diagonal line in Fig.~\ref{fig:lambda-2D}.}
 \label{fig:phys-upto-L20}
\end{figure}

Figure~\ref{fig:phys-upto-L20} shows system-size dependences of $D$ [panel (a)] and $S(\bm{Q})/\Ns$ [(b)] up to $L$=12, obtained by the mVMC calculation.
We found that these physical quantities are well converged for $L\ge 6$. 
For all the system sizes, the Mott transition is characterized by a level crossing between the AFI and PMM states, and a finite jump in the properties, indicating their first-order nature of the transitions.

\begin{figure*}[!t]
 \centering
 \resizebox{0.8\textwidth}{!}{\includegraphics{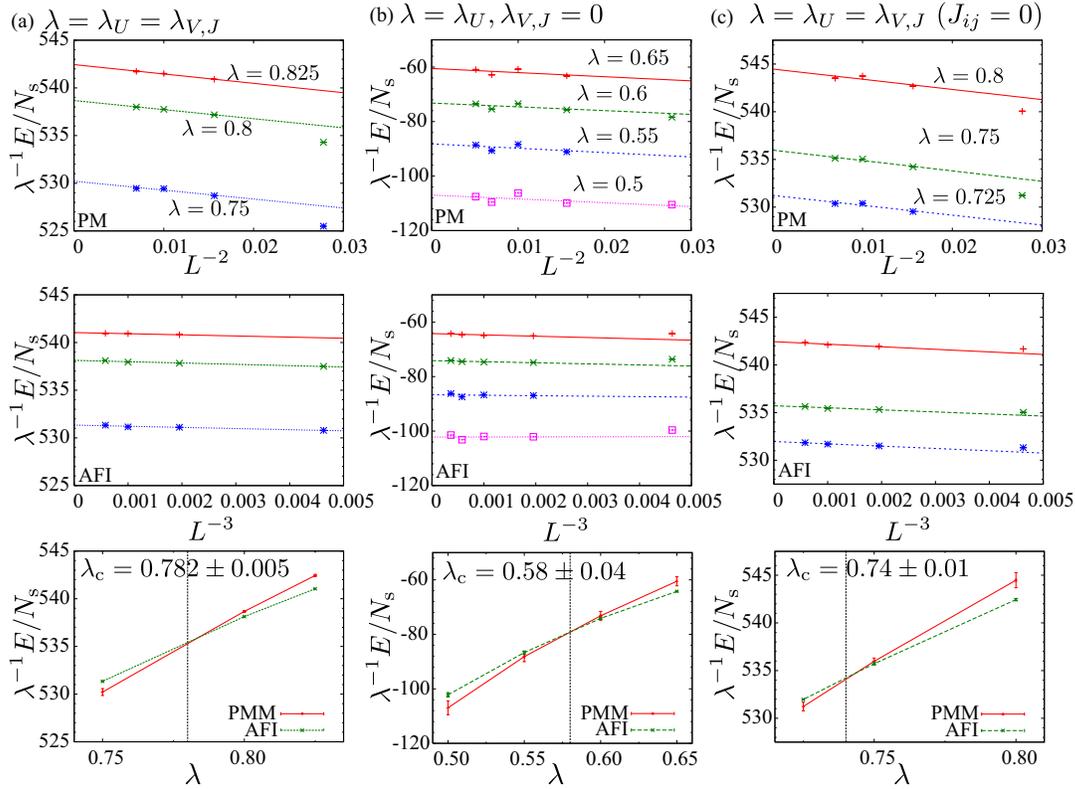}}
 \caption{(Color online) Extrapolation of energy $E/\Ns$ to the thermodynamic limit around the Mott transition along (a) diagonal line of $\lU=\lV$ in the phase diagram (Fig.~\ref{fig:lambda-2D}), (b) horizontal line of $\lV=0$, and (c) diagonal line of $\lU=\lV$ with $J_{ij}=0$. The energy $E$ is given in the unit of meV. The top and middle panels illustrate the size extrapolations of the PMM and AFI, respectively.  The bottom panels show energy comparisons in the thermodynamic limit for the PMM and AFI phases.} 
 \label{fig:E-extrapolation}
\end{figure*}

We next consider the ground-state properties in the thermodynamic (bulk) limit. To this end we extrapolate the energies per site $E/\Ns$ of the AFI and PMM states around $\lc$.
For the AFI state, we employ the scaling form $\Delta E/\Ns \propto L^{-3}$ given by the spin-wave theory for the two-dimensional quantum Heisenberg antiferromagnets~\cite{Huse88} with $\Delta E$ being the finite-size correction of the total energy. For the PMM state, we extrapolate the energy by following $\Delta E/\Ns\propto L^{-2}$.~\cite{comment1}
The top and middle panels in Fig.~\ref{fig:E-extrapolation}(a) illustrate the procedure of the size extrapolation for PMM and AFI, respectively.
The critical interaction ratio was estimated as $\lc=0.782\pm0.005$ [the bottom in Fig.~\ref{fig:E-extrapolation}(a)] as the crossing point of the two energies extrapolated to the thrmodynamic limit.

After determining $\lc$, we calculated the ground-state physical quantities in the thermodynamic limit. 
The doublon density $D$ in the thermodynamic limit for each locally stable state was estimated with the scaling form $D(L=\infty)-D(L)\propto L^{-1}$.
We also calculated the staggered magnetization $\ms$ by extrapolating $S(\bm{Q})/\Ns$ to the thermodynamic limit as $\ms^2-S(\bm{Q},L)/\Ns\propto L^{-1}$.
This scaling form for $\ms^2$ is suggested by the spin-wave theory for the two-dimensional quantum Heisenberg antiferromagnets~\cite{Huse88}. The extrapolated $D$, $\ms^2$, and $\ms$ are shown in Fig.~\ref{fig:phys-upto-L20}(c). Finite jumps in $D$ and $m_\mathrm{s}$ in the thermodynamic limit indicates the first-order nature of the Mott transition, which is consistent with the experimental behavior observed in the pressure-controlled Mott transition of $\kappa$-Cl~\cite{Kagawa}. The obtained ordered moment is $m_s\sim 0.22$ near the metal-insulator transition. This value is much smaller than that for the spin-1/2 Heisenberg model on the square lattice ($\sim 0.3$),~\cite{Reger88} while it is comparable to that in the antiferromagnetic phase near the Mott transition in the geometrically frustrated lattice.~\cite{Kashima,Morita,Mizusaki}

\begin{figure}[p]
 \centering
\resizebox{0.425\textwidth}{!}{\includegraphics{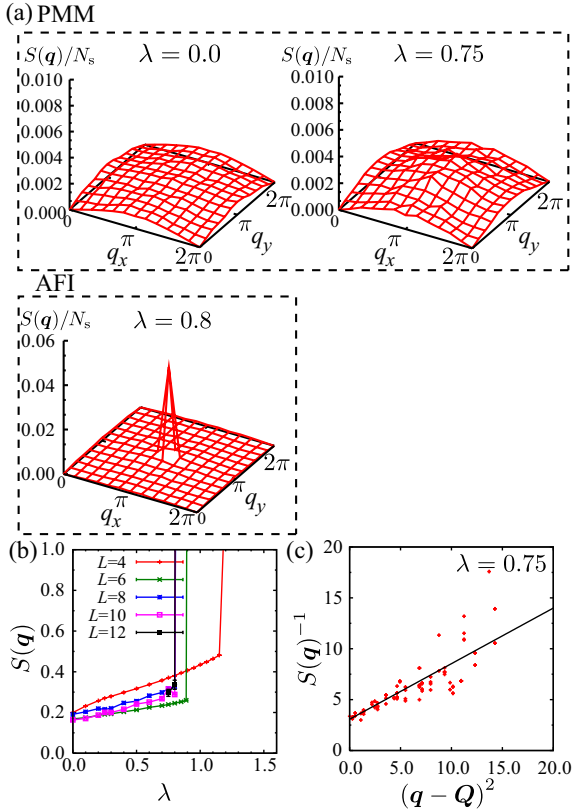}}
\caption{(Color online) (a) Momentum-resolved $S(\bm{q})$ at $\lambda=0, 0.75$, and 0.8 for $L=12$. 
(b) $\lambda$ dependence of $S(\bm{Q})$. (c) Ornstein-Zernike type fit of $S(\bm{q})$ at $\lambda=0.75$ (see text).
}
 \label{fig:sq-L12}
\end{figure}

In Fig.~\ref{fig:sq-L12}(a), we plot the momentum-resolved spin structure factor $S(\bm{q})$ calculated for $L=12$ near the metal-insulator transition.
There is no essential difference between the result for $\lambda=0$ and that for $\lambda=0.75$, indicating the absence of the antiferromagnetic spin fluctuations near the metal-insulator transition.
This is also seen more directly in the absence of system-size dependence for $S(\bm{Q})$ [Fig.~\ref{fig:sq-L12}(b)]. 
The antiferromagnetic correlation length $\xi_\mathrm{AF}$ was estimated from a fitting of the Ornstein-Zernike form 
\begin{align}
	S(\bm{q})&=\frac{S(\bm{Q})}{1+\xi_\mathrm{AF}^2(\bm{q}-\bm{Q})^2} \label{eq:Orstein}
\end{align}
to have $\xi_\mathrm{AF}/a=0.42\pm 0.02$ with $a$ being the lattice spacing in terms of the square lattice [see Fig.~\ref{fig:sq-L12}(c)].
This is a strong indication that the antiferromagnetic correlation does not develop even near the metal-insulator transition, 
which is consistent with the experimental observation of $\kappa$-NCS at low $T$; i.e., $T<55$ K~\cite{Kawamoto}.
We note that the enhancement of the antiferromagnetic spin correlations above 55 K in $\kappa$-NCS may be due to proximity effects of the Mott transition.
For instance, at $\lambda=0.75$ (about 4\% below $\lc$, which may be a typical plausible value for the real compound), the energy difference per site between 
AFI and PMM is on the order of 1 meV $\simeq$ 10 K, and thus 
it is likely that the antiferromagnetic metastable state partially contributes to the finite-$T$ manifold and enhances the antiferromagnetic correlations. 
Further studies for finite-$T$ properties are desirable for this topic.

\begin{figure}[t]
 \centering
 \resizebox{0.4\textwidth}{!}{\includegraphics{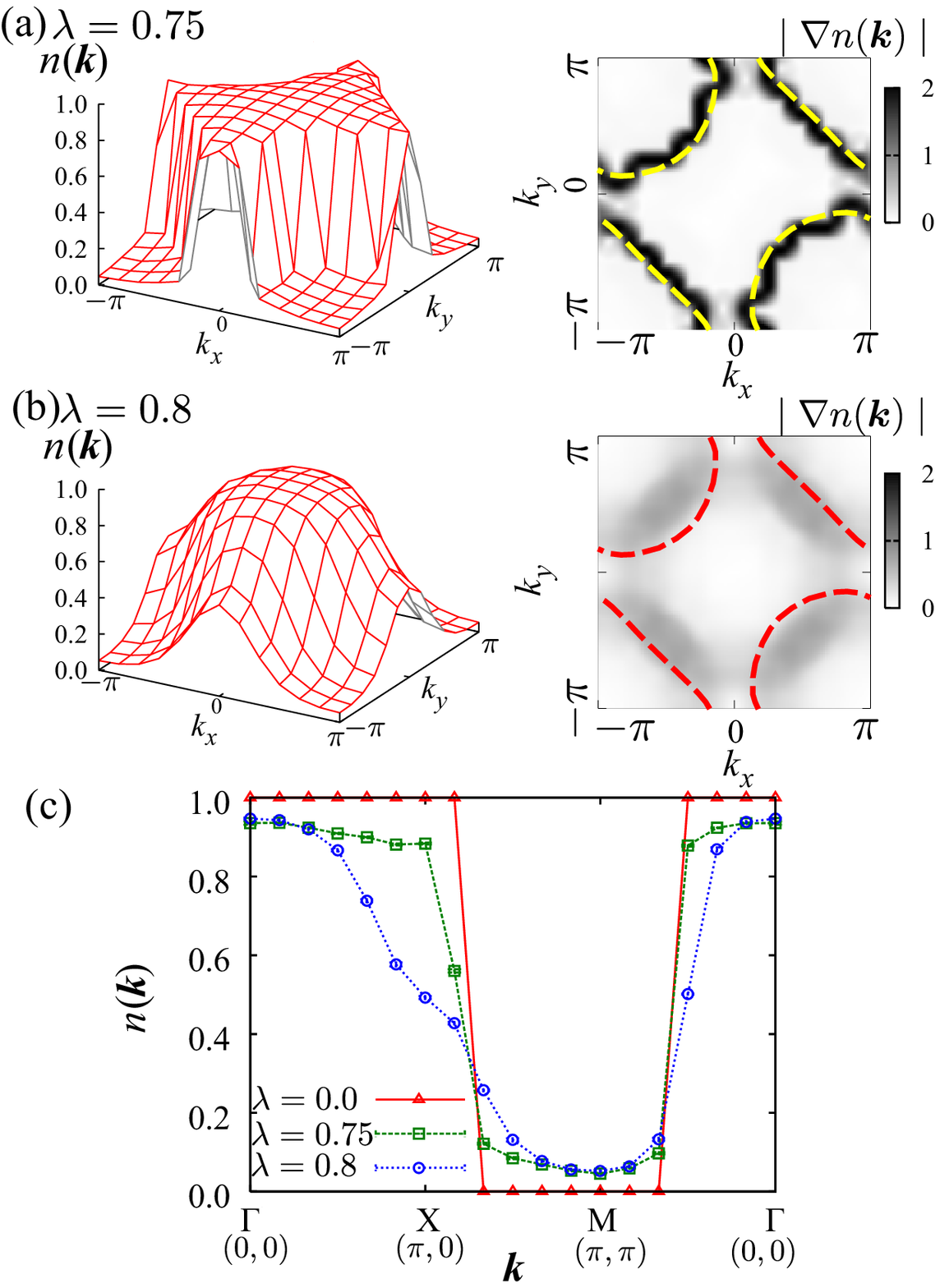}}
 \caption{(Color online) Momentum distribution $n(\bm{k})$ and contour plots of $|\nabla n(\bm{k})|$ at $\lambda$=0.75 (a) and 0.8 (b), respectively ($L=12$). Here, $n(\bm{k})$ is calculated by using the value interpolated by the bi-cubic interpolation of $n(\bm{k})$. The broken lines denote the Fermi surface at $\lambda=0$ and $L=\infty$. (c) $\lambda$ dependence of $n(\bm{k})$ along symmetry lines.} 
 \label{fig:momentum-L20}
\end{figure}

A recent mVMC study indicates that the charge excitation gap is partially formed 
in the metallic phase as a precursor of the Mott gap for the square-lattice Hubbard model with next-nearest hopping~\cite{Tahara2}, where an ``arc-like'' Fermi surface 
is observed near the Mott transition, i.e., $U_\mathrm{c}/t=3.3$ at $t^\prime/t=-0.3$.
To make a comparison with this result, we plot in Fig.~\ref{fig:momentum-L20} the momentum distribution $n(\bm{k})$ and its gradient $|\nabla n(\bm{k})|$ for $\lambda$=0.75 [panel (a)] and 0.8 [(b)].
The Fermi surface of $\lambda=0$ is denoted by dotted lines in the contour plot. 
The electron correlations smear the jumps in $n(\bm{k})$ and this effect is more significant around $(\pi,0)$ than around $(\pi/2,\pi/2)$ [Fig.~\ref{fig:momentum-L20}(c)].
We remark that the present result has no ``arc-like'' structure in the $|\nabla n(\bm{k})|$ plot near the Mott transition in contrast to the previous study. This may be due to the large critical interaction strength, i.e, $U_\mathrm{c}\simeq 7.5$, and the resultant strong first-order nature.~\cite{comment2}

To discuss effects of the off-site interaction $V_{ij}$ and $J_{ij}$, we compare the critical interaction ratio $\lc$ obtained with/without these parameters. By switching off $V_{ij}$ and $J_{ij}$ (i.e., the line along $\lV$=0 in Fig.~\ref{fig:lambda-2D}), we obtained $\lc$=0.58$\pm$0.04 [Fig.~\ref{fig:E-extrapolation}(b)]. By switching off only $J_{ij}$ (i.e., the $\lU$=$\lV$ line but $J_{ij}$=0), we obtained $\lc$=0.74$\pm$0.01 [Fig.~\ref{fig:E-extrapolation}(c)].  Since $\lc$ is 0.782$\pm$0.005 for the model having $V_{ij}$ and $J_{ij}$, we have an appreciable enhancement of $\lc$ with $V_{ij}$ and a minor modification by $J_{ij}$. The mechanism of the increase in $\lambda_c$ by $V_{ij}$ is as follows: With introducing $V_{ij}$, the creation energy for a  
doublon-holon pair in the Mott insulator is reduced from $U$=0.64 eV to $U-V_{ij}$=0.45-0.48 eV. 
Furthermore, $V_{ij}$ also reduces the local moment $m_{\rm local}=\sqrt{1-2D}/2$ by increasing the doublon density.
The introduction of $J_{ij}$, on the other hand, favors a ferromagnetic correlation on the nearest-neighboring bonds, thus destabilizes the antiferromagnetic solution.

\subsection{Charge-ordered phase} \label{Charge-ordered insulator}
In general, off-site Coulomb interactions tend to stabilize charge-ordered states.
This effect is not fully considered in the previous section, in which the analyses are limited to $2\times 2$ orders;
e.g., at half filling, the nearest-neighbor repulsion stabilizes three-fold charge orders.
In this section, we examine the possibility of a charge ordering induced by $V_{ij}$ in more extended parameter space with a more generalized variational wave function.

For calculations along the lines of $\lV=1$ and $\lV=0.7$ in the phase diagram of Fig.~\ref{fig:lambda-2D}, 
we extend the pairing wave function $|\phi_{\rm pair}\rangle$ by allowing $f_{ij}$ to have a $6\times 2$ sublattice structure (see Fig.~\ref{fig:sblattice-6x2}).
This allows us to search magnetic and charge orders with ordering vectors of $(\pi,\pi)$ and $(2\pi/3,2\pi/3)$ on equal footing.
To identify charge orders, we calculate the charge structure factor
\begin{eqnarray}
	N(\bm{q}) &=& \frac{1}{\Ns} \sum_{ij}(n_i-1)(n_j-1) e^{-{\rm i}\bm{q}\cdot (\bm{r}_i-\bm{r}_j)}.
\end{eqnarray}

\begin{figure*}[!p]
 \centering
 \resizebox{0.5\textwidth}{!}{\includegraphics{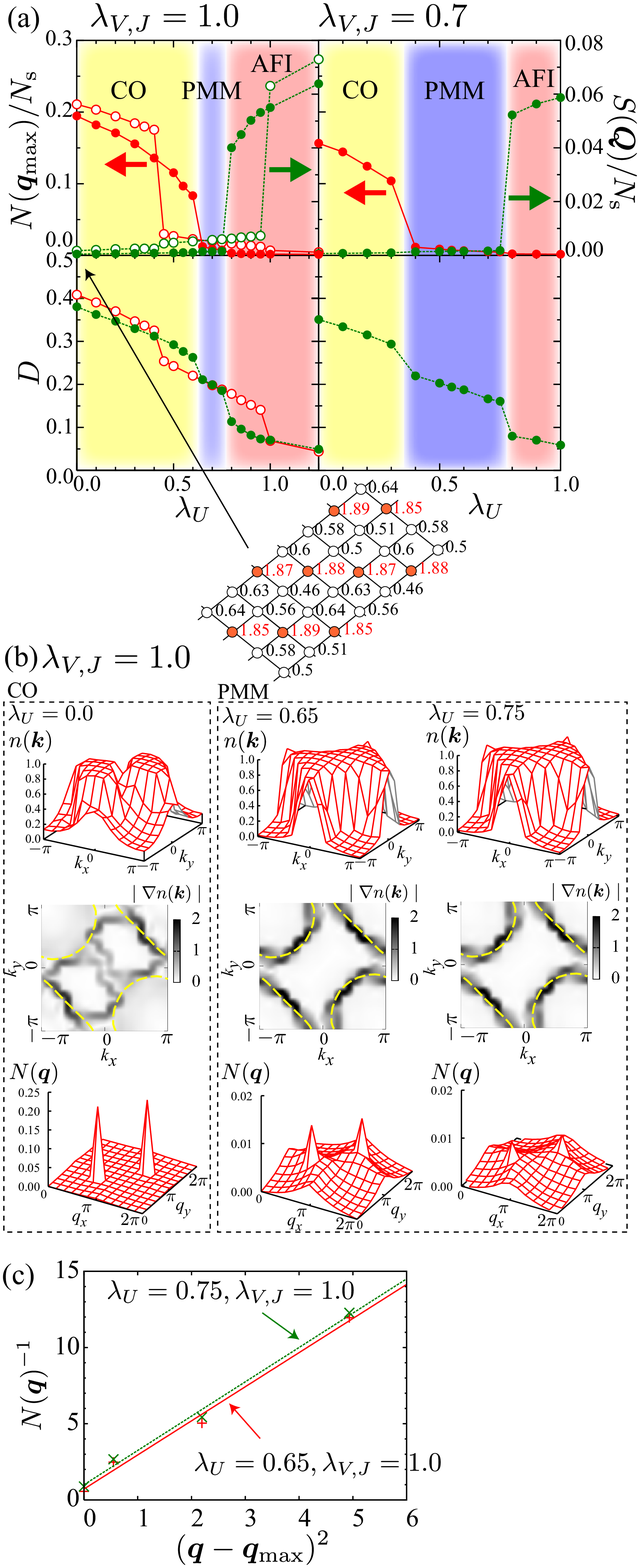}}
 \caption{
 (Color online) (a) Calculated $N(\bm{q}_\mathrm{max})$, $S(\bm{Q})$, and $D$ at $\lV=1.0$ and 0.7 as functions of $\lU$. $N(\bm{q}_\mathrm{max})$ is the peak value of the charge structure factor $N(\bm{q})$. The open and filled circles denote data for $L=6$ and $12$, respectively. For all $\lU$ and $\lV$, we found that $\bm{q}_\mathrm{max}=(2\pi/3,2\pi/3)$, and equivalently at $(4\pi/3,4\pi/3)$. 
The bottom distorted square lattice denotes the charge density $\langle n_{i} \rangle $ calculated at $\lU=0.0$ and $\lV=1.0$ for system size $L=12$.
 (b) Momentum-resolved $n(\bm{k})$, $|\nabla n(\bm{k})|$ and $N(\bm{q})$ at $\lV=1.0$ and $L=12$.
 The broken (yellow) lines denote the Fermi surface at $\lU=\lV=0$ and $L=\infty$.
 (c) Ornstein-Zernike type fit for $N(\bm{q})$ along the $(0,0)$-$(2\pi/3,2\pi/3)$ line. The system size is $L=12$.
}
 \label{fig:CO}
\end{figure*}

The $L$=6 and $L$=12 results calculated for $\lV=1.0$ and $0.7$ are shown in Fig.~\ref{fig:CO}.
By decreasing $\lU$ with $\lV$ fixed, the system turns into the CO from the PMM.
At $\lV=1.0$ and 0.7, this transition is found to be of the first order as seen in the panel~(a).
The CO phase is characterized by a three-fold (\textit{rich-poor-poor}) order as seen in the $\langle n_i \rangle$ profile [the bottom of the panel (a)] calculated at $\lU=0.0$ and $\lV=1.0$.
The left side of the panel~(b) displays our calculated $n(\bm{k})$, $|\nabla n(\bm{k})|$, and $N(\bm{q})$ at $\lU$=0.0 and $\lV$=1.0 for $L=12$. We found sharp peaks in $N(\bm{q})$ at $(2\pi/3, 2\pi/3)$ and $(4\pi/3, 4\pi/3)$, due to the three-fold CO order.
A UHF calculation indicates that the CO state has a small charge gap of about $0.2$~eV associated with a two-fold bond order at $\lU$=0.0 and $\lV$=1.0 for $L$=24 (not shown).
However, we could not conclude whether or not the CO phase is insulating in the mVMC calculations [see Fig.~\ref{fig:CO}(b)],
because the system sizes in the present study are not sufficiently large for the size extrapolation of the charge gap.

A remarkable observation here is that the peaks in $N(\bm{q})$ exist even in the PMM phase as seen in the right side of Fig.~\ref{fig:CO}(b).
We estimate a correlation length of the charge fluctuations $\xi_\mathrm{c}$ in terms of the Ornstein-Zernike function
\begin{align}
	N(\bm{q})&=\frac{N(\bm{q}_\mathrm{max})}{1+\xi_\mathrm{c}^2(\bm{q}-\bm{q}_\mathrm{max})^2} \label{eq:Orstein2}
\end{align}
with $\bm{q}_\mathrm{max}=(2/3\pi,2/3\pi)$.
The result is shown in Fig.~\ref{fig:CO}(c): We obtained $\xi_\mathrm{c}/a$=1.8$\pm$0.6 and 1.5$\pm$0.3 for $\lU$=0.65 and $\lU$=0.75 (PMM), respectively, at $\lV=1.0$.
These correlation lengths are appreciably longer than those of the antiferromagnetic correlations;  
we obtained $\xi_\mathrm{AF}/a$=0.33$\pm$0.01 and 0.34$\pm$0.01 for $\lU$=0.65 and 0.75, respectively.
These indicate that the charge fluctuations are more dominant in the PMM phase than the antiferromagnetic correlations.
More interestingly, in the PMM phase at $\lambda_{V,J}=1.0$ and $\lambda_U=0.75$, 
the Fermi surface becomes smeared around $(\pm\pi,0)$ and $(0,\pm\pi)$, and as a consequence, 
``Fermi arcs'' appear around $(\pm\pi/2,\pm\pi/2)$ [see the right panels of Fig.~\ref{fig:CO}(b)].
This is in sharp contrast to Figs.~\ref{fig:momentum-L20}(a), where we did not see the ``arc-like" structure in $|\nabla n({\bm k})|$.
This indicates that the emergence of the arc-like structure is ascribed to the enhanced charge fluctuations. 
Recently, Fermi-arc like behavior has been observed in a metallic phase adjacent to a CO phase for
a metallic layered	nickelate Eu$_{2-x}$Sr$_x$NiO$_4$~\cite{Uchida11}.
The formation of a Fermi arc might be characteristic to metallic states with strong charge fluctuations.

The results are summarized in Fig.~\ref{fig:lambda-2D} as the $\lU$-$\lV$ phase diagram. The PMM is sandwiched by CO and AFI and 40\% reduction of $U$ in PMM realized in $\lV$=1 causes the transition to CO. The CO transition line and AF transition line merge in the strong correlation limit, i.e., $\lU,\lV\rightarrow +\infty$. The critical interaction ratio in this limit is given by $\lU/\lV =\frac{U/4}{V+V^\prime/2-V^{\prime\prime}}\simeq 0.73$ based on a simple estimate of the ground-state energy 
(see Appendix).

\section{Summary and discussion} \label{Summary}
We have performed mVMC calculations for an {\it ab initio} low-energy effective model of  $\kappa$-NCS.
The ground state of this compound has turned out to be close to the Mott transition.
Within 20\% change in the {\it ab initio} interaction parameters, it undergoes a transition between an antiferromagnetic insulator and a metal.  The real compound is known to be scarcely metallic at low temperatures while semiconducting above 90K with enhanced antiferromagnetic correlations. The calculated result reproduces these basic experimental results.
In this first challenge of the {\it ab initio} calculation of $\kappa$-NCS, 
the present result has proven a good accuracy of the three-stage {\it ab initio} scheme even for strongly correlated and complex organic compounds.
  
However, strictly speaking, the real compound of $\kappa$-NCS becomes metallic at low temperatures implying that the present result shows approximately 20\% 
underestimate of the transition parameter in terms of the interaction strength. A possible origin of this discrepancy may be the overestimate of the antiferromagnetic region by the mVMC method in the strong correlation regime.  Actually, the benchmark of the metal-insulator transition for the 
Hubbard model on the square lattice with the next-nearest-neighbor 
transfer $t'=0.2\sim 0.3$ has shown nearly 10\% overestimate of the antiferromagnetic insulating 
phase as compared to a more precise PIRG result (the metal-insulator boundary is around $U/t=3.6$ by the PIRG estimate against the mVMC result $U/t\sim 3.3$).\cite{Tahara,Kashima}  
Therefore, in more strongly correlated region as in this case in comparison to the benchmark, it is likely to overestimate the stability of the antiferromagnetic insulator as well in a similar rate; 
in fact, 20\% overestimate has been confirmed in the present comparison with the exact diagonalization for a small cluster ($L$=4).
The development of a more accurate low-energy solver is a future important issue.
In the PIRG, it is possible to estimate the boundary without an explicit bias beyond the present mVMC and it is worthwhile to do so, although the present parameter with a large frustration and interaction may demand a heavy calculation. 

Another possible origin of the overestimate of the stability of the antiferromagnetic
insulator is dynamical effects of the HOMO bonding band eliminated in the present downfolding procedure. If the dynamical polarization arising from the HOMO bonding band becomes important, one has to consider the two-band model as the low-energy effective model. Charge fluctuations within the dimer of the BEDT-TTF molecule may melt the antiferromagnetic order and shift the Mott transition boundary to a higher $U$ value. 
This may be related to the dielectric anomaly recently observed in the experiments around 6K of $\kappa$-CN~\cite{Abdel-Jawad10, Manna} as discussed below.

The insulating side of this compound is shown to have a remarkably suppressed ordered moment ($\sim 0.22$) of the antiferromagnetic order. This is clearly ascribed to the geometrical frustration effect as well as the proximity of metals.  A quantum spin liquid is stabilized for more frustrated case as $\kappa$-CN,
and this compound is also barely antiferromagnetic, suggesting that the ordered moment for $X=$ Cu[N(CN)$_2$]Cl (a very similar compound but just in the insulating side) is likely to have a small ordered moment similar to the present estimate. 

By applying {\it ab initio} downfolding scheme to $\kappa$-NCS, we have shown that the nearest-neighbor and next-nearest-neighbor off-site Coulomb interactions $V$ of the 2D low-energy effective model are unexpectedly large, i.e., $V/U\sim 1/4$. 
These large off-site Coulomb interactions stabilize the paramagnetic metal through the excitonic effect.

Experimentally, $\kappa$-NCS is known to be superconducting at low temperatures. 
In this paper, we have assumed only the normal state in the metallic 
phase and have not examined the possibility of superconductivity in 
detail to focus on the metal-insulator boundary. 
The phase competition involving the superconducting state is 
left for future study, because it requires very subtle comparison of the stability of the superconducting state.

It has been proven that the three-stage scheme is powerful 
for the {\it ab initio} calculation of the organic conductors. 
It is an intriguing issue to study the nature of the 
quantum spin liquid for $\kappa$-CN as well as the unconventional Mott transition for $\kappa$-Cl by using the present accurate {\it ab initio} method.        

Before closing this paper, we make a brief discussion on a recent experimental observation of a relaxer-like dielectric anomaly for dimer-Mott insulator $\kappa$-CN~\cite{Abdel-Jawad10,Manna}, which is known as a spin-liquid material.
As its possible origin, effects of charge fluctuations within the ET dimers have been extensively investigated theoretically~\cite{Abdel-Jawad10, Naka10, Hotta09}.
These effects are not taken into account in the present study in which the {\it ab initio} model is constructed based on the dimer basis.
However, the present study has revealed that charge fluctuations are also enhanced by the inter-dimer Coulomb interactions in a wide range of the parameter space, especially near the charge-ordered phase, which is not very far from the realistic choice of the parameters $\lambda_U=\lambda_V=1$.
This indicates that the off-site (inter-dimer) Coulomb interactions may also play a certain role in the dielectric anomaly observed in $\kappa$-CN.
First-principles studies on $\kappa$-CN are left for future study.
In particular, it is of great interest to understand effects of inter/intra-dimer charge fluctuations based on an {\it ab initio} two-band model of $\kappa$-CN.

\begin{acknowledgments}
	The authors thank Daisuke Tahara for the use of his mVMC code and useful comments. Numerical calculation was partly carried out at the Supercomputer Center, Institute for Solid State Physics, Univ. of Tokyo. This work was supported by Grant-in-Aid for Scientific Research (No. 22740215, 22104010, 22340090, and 23110708), from MEXT, Japan. A part of this research has been funded by MEXT HPCI Strategic Programs for Innovative Research (SPIRE) and Computational Materials Science Initiative (CMSI).	
\end{acknowledgments}

\appendix
\section{Critical interaction ratio of the three-fold charge-order transition in the classical limit}\label{sec:appedix1}
We assume that the ground state is non-magnetic and has a three-fold charge order.
In the limit of $\lU,\lV\to +\infty$ (classical limit), the energy per site $E/\Ns$ is given by
\begin{eqnarray}
&&E/\Ns\\ \nonumber
&=& \lU E_U + \lV E_V + \lV E_{V^\prime} + \lV E_{V^{\prime\prime}},\\
&=&\frac{1}{3}\left\{\lU\frac{U}{4} -\lV\left(V+\frac{V^\prime}{2}-V^{\prime\prime}\right)\right\}\times \\ \nonumber
&&(n_1^2+n_2^2+n_3^2)+\mathrm{const.}
\end{eqnarray}
where
\begin{eqnarray}
	E_U&=&\frac{U}{12}(n_1^2+n_2^2+n_3^2),\\
	E_V&=&\frac{V}{3}\left\{9n^2-(n_1^2+n_2^2+n_3^2)\right\},\\
	E_{V^\prime}&=&\frac{V^\prime}{6}\left\{9n^2-(n_1^2+n_2^2+n_3^2)\right\},\\
	E_{V^{\prime\prime}}&=&\frac{V^{\prime\prime}}{3}(n_1^2+n_2^2+n_3^2).
\end{eqnarray}
Here, $n_1$, $n_2$ and $n_3$ are the site occupancies for the three sublattices, and the mean site occupancy is given by $n\equiv(n_1+n_2+n_3)/3=1$.
The Cauchy-Schwarz inequality tells us that $n_1^2+n_2^2+n_3^2$ has it minimum $1/3$ when $n_1=n_2=n_3=1/3$.
This indicates that a three-fold charged-ordered state becomes more stable than the uniform state when $\lU\frac{U}{4} -\lV\left(V+\frac{V^\prime}{2}-V^{\prime\prime}\right)<0$.
Therefore, the critical interaction ratio of the three-fold charge-order transition is given by
\begin{eqnarray}
	\lU/\lV=\frac{\frac{U}{4}}{V+\frac{V^\prime}{2}-V^{\prime\prime}}
\end{eqnarray}
in the classical limit.


\begin{thebibliography}{27}
\bibitem{comment1} This finite-size correction is expected from the bulk contribution of the error in the discrete $k$ summation. We confirmed that, in the absence of electron correlation, the system-size dependence of the energy indeed follows this scaling in the case of the square lattice with the periodic-antiperiodic boundary condition.
\bibitem{comment2} The first-order nature of the Mott transition tends to become stronger as $U_\mathrm{c}$ increases because the jump in $\ms$ becomes larger.
\end{thebibliography}
\end{document}